\documentclass[showpacs,twocolumn,nofootinbib,
floatfix]{revtex4-1}
\usepackage{amsfonts,amsmath,amssymb,amsthm}
\usepackage[bookmarks, bookmarksopen]{hyperref}
\usepackage{graphicx}
\usepackage{color}
\usepackage[utf8]{inputenc}
\usepackage[T1]{fontenc}
\usepackage{lmodern}
\usepackage[right]{rotating}
\usepackage{longtable}
\usepackage{array}
\usepackage{multirow}
\usepackage{paralist}
\usepackage{siunitx}
\usepackage{physics}

\begin{document}
\title{Spherically symmetric configurations of General Relativity in presence of scalar field: separation of test body circular orbits}
\date{\today}

\author{O.~S.~\surname{Stashko}}
\author{V.~I.~\surname{Zhdanov} }\thanks{e-mail: valeryzhdanov@gmail.com}
\affiliation{Taras Shevchenko National University of Kyiv,\\ Astronomical Observatory, Observatorna st., 3, Kiev, 04053 Ukraine}

\begin{abstract}
We study the test body circular orbits in a gravitational field of a static spherically symmetric  object in presence of a minimally coupled nonlinear scalar field. We generated a two-parametric family of massless scalar field potentials and corresponding solutions of Einstein -- scalar field equations in an analytic form. The results are presented by means of hypergeometric functions; they describe either a naked singularity (NS) or a black hole (BH). We show that for the solutions found there is always an unbounded region of stable circular orbits (SCO), which is qualitatively analogous to the Schwarzschild BH case. Besides that, for a considerable range of the angular momenta and parameters of the family, there can exist  (as for the case of NS and BH) the other SCO region, which is separated from the previous one by a ring of unstable circular orbits.  We present another solutions showing that analogous distribution of SCO can exist in case of a massive scalar field, also for  BH and NS cases.
\end{abstract}

\LTcapwidth=\columnwidth


\maketitle{}

\section{Introduction}
\label{sec:intro}

Scalar field theories occupy a considerable sector in gravitational physics \cite{Will,Berti}, especially in cosmology where additional  fields are extensively used to study the inflationary era of early Universe, in dynamical models of dark energy etc (see, e.g., \cite{Plank, Linde, Novosyadlyi} for a review and references therein). There is no direct evidence that, at present epoch, the scalar fields play some role in astrophysical systems \cite{Will} and one might think that  effects of the cosmological fields must be negligible. Nevertheless, since the early works \cite{Fisher, Janis}, it is well known that any (arbitrarily small!) presence of a scalar field can cause a cardinal change of the space-time topology in the vicinity of a compact object. However, this as itself does not mean the existence of significant observational effects.

The search for a smoking gun of scalar fields as well as effects of alternative gravitation theories requires investigation of the geodesic structure associated with solutions of the Einstein-scalar field equations and their modifications. The very first steps involve of the test body orbits around gravitating center (see, e.g., \cite{Solovyev,Chowdhury,Vieira,Stuchlik,Pugliese,Pugliese2,Pugliese3,Boshkayev} and references therein). It is well known that stable circular orbits (SCO) in Schwarzschild and Kerr black hole space-times of General Relativity form a connected structure without gaps. A number of examples show that this picture can be violated  in presence of a naked singularity (NS); discontinuous structures of SCO can arise  in case of charged and uncharged objects of General Relativity \cite{Pugliese,Pugliese2,Pugliese3},  in case of a quadrupole source \cite{Boshkayev}. In presence of scalar field, there also can exist disconnected regions  of SCO separated by a ring of unstable orbits \cite{Chowdhury}. The result \cite{Chowdhury} concerns  spherically symmetric space-times (with NS) and linear massless scalar fields . The questions arise, is NS necessary for the existence of such a discontinuous structure? Can these structures occur in scalar field black hole space-times?

The aim of our paper is to present examples with a non-trivial self-interaction potential $V(\phi)$ of the scalar field $\phi$ showing that

this disconnected structure of SCO can exist in the black hole (BH) space-time, not only in presence of NS, and

the potential $V(\phi)$ can describe either massless or massive scalar field.

 Note that our black hole solutions discussed below involve  potentials that are negative in some regions, so they do not contradict to the no-scalar-hair theorem \cite{Bekenstein}.

The existence of the discontinuous structures is very important as it may have relevance to  accretion disks around real astrophysical  objects.  Common ideas about stellar mass BHs and supermassive BHs in active galactic nuclei deal with a circular motion of a surrounding matter. Obviously, the  consideration of an accretion disk does not reduce only to an analysis of  geodesics  and a lot of complicating factors must be taken into consideration: the ionized gas pressure, turbulence, magnetic fields, radiation etc (see, e.g. \cite{Novikov-Thorne, Reynolds-Nowak, Lasota}). However, the presence of an unstable orbits region, separating  the regions of stable ones, can hardly be changed by these factors. It is difficult to find reasons, which could destroy this separation, leading to a gap in the accretion disk,  even in presence of an intricate physics.

Properties of inner parts of accretion disks in the Galactic and extragalactic systems are studied by means of the X-rays observations, including  investigation of fluorescent iron lines in the X-ray spectra  \cite{Guilbert,Lightman,Fabian, Reynolds-Nowak}. If the above  gap in the accretion disk  really exists,  this could  affect the form of the lines. This may be of interest for testing the relativistic gravitation theories.

In the present paper we study test-body orbits in static spherically symmetric asymptotically flat space-times of General Relativity in presence a non-linear scalar field minimally coupled with gravity.  In Section \ref{orbits} we derive general conditions for the appearance disconnected structure of SCO. In Section \ref{sec:setup} we found a family of special solutions to Einstein - scalar field equations,  which describe configurations with a positive mass. With this aim we  use a known method (see \cite{Bronnikov, BronShikin, BronnikovFabris, Nikonov, Azreg2009}) of generation of special spherically symmetric solutions to these equations along with corresponding potentials. In this section we deal with  massless scalar fields. Investigation shows the occurrence of the disconnected distribution of SCO, both in case of BH and/or NS, for certain domain of the family parameters. At the same time, the other choice of the parameters ensures the absence of discontinuities in the distribution. In Section \ref{numerical} we show limits on the family parameters which allow or prohibit the disconnected structure of SCO in different situations.
Section \ref{generalizations} deals with some generalizations including  massive  scalar field. The results are summarized up in Section \ref{conclusions}.

\section{Basic relations and notations}
\label{basic}
The metric of a static spherically symmetric space-time can be written as
\begin{equation}
\label{eq2.0}
ds^2 = A(x)dt^2 - {B(x)}dx^2 - r^2(x)dO^2 ,
\end{equation}
where  $dO^2=d\theta ^2 + \sin^2(\theta)d\varphi^2$ and $A>0,\,B>0$ in a  static region.  In case of the usual Schwarzschild coordinates one can choose $r$ as the radial variable yielding $ds^2=Adt^2-B(dx/dr)^2dr^2-r^2 dO^2$. In this Section we use essentially some of the results of   \cite{Bronnikov, BronnikovFabris, Azreg2009}. Following \cite{Bronnikov, BronnikovFabris, Azreg2009} we use  the other choice of coordinates by putting $B=1/A$:
\begin{equation}
\label{eq2}
ds^2 = A(x)dt^2 - \frac{dx^2}{A(x)} - r^2(x)dO^2,
\end{equation}
and we say that $x_0$ is a point of center if $r(x_0)=0$ and $r(x)>0$ for $x>x_0$.

We assume   $A(x),\,r(x)\in C^{(2)}$, $\phi\in C^{(1)}$,
\begin{equation}
\label{asy_flatness}
 r(x)=x+o(1/x),\quad  r'(x)=1+o(1/x),
\end{equation}
and
\begin{equation}
\label{asy_flatness++}
 A(x)=1-2m/x+o(1/x),\, m> 0,
\end{equation}
as $x\to \infty$ specifying an asymptotically flat space-time   of an isolated system with positive total mass $m$. For the scalar field we assume
$\lim\limits_{x\to \infty}\phi(x)= 0$.

The Einstein equations in presence of a self-interacting minimally coupled  scalar field $\phi$ can be derived from the action functional
\begin{equation}
	\label{eq1}
	S = S_{GR} + \int {d^4} x\sqrt {\vert g\vert } \left[ {g^{\mu \nu
		}\phi _{,\mu } \phi _{,\nu } - 2 V(\phi )} \right] \quad ,
\end{equation}
\noindent
where $S_{GR} $ is the standard gravitational action of the General Relativity ($c=8\pi G=1$), and $V(\phi )$ is a self-interaction potential to be specified below. In case of metric  \ref{eq2}  this
 yields the following system
\begin{equation}
	\label{eq3}
	\frac{d}{dx}\left( {\frac{dA}{dx}r^2} \right) = - 2r^2V(\phi ),
\end{equation}
\begin{equation}
		\label{eq3b}
	\frac{d^2r}{dx^2} + \frac{1}{2}r\left( {\frac{d\phi }{dx}} \right)^2 = 0 ,
\end{equation}
\begin{equation}
		\label{eq3c}
	A\frac{d^2r^2}{dx^2} - r^2\frac{d^2A}{dx^2} = 2 .
\end{equation}
Variation of  (\ref{eq1}) with respect to $\phi$ yields one more equation, which is not independent from equations (\ref{eq3}---\ref{eq3c}); therefore, we do not write it.

From equation (\ref{eq3b} it follows
\begin{equation}
\label{asy_flatness+}
\frac{d^2r}{dx^2}\le 0,\quad \frac{dr}{dx}\ge 1  ,
\end{equation}
whence the point of center $x_0$ is a simple root.

Equation (\ref{eq3c}) can be written as
\begin{equation*}
\frac{d}{dx}\left[r^4	\frac{d}{dx} \left(\frac{A}{r^2}\right)\right] = -2 ;
\end{equation*}
on account of   (\ref{asy_flatness}) this yields
\begin{equation}
	\label{eq4}
	A(x) = r^2(x)\int\limits_x^\infty {\frac{2{x}' - C}{r^4({x}')}} d{x}'
	\quad 	,
\end{equation}
\noindent
where  $C$ is an integration constant.

Under conditions (\ref{asy_flatness}, \ref{asy_flatness++}),  we derive $C=6m$.

In view of (\ref{asy_flatness}, \ref{eq3}, \ref{eq3b}) we get
\begin{equation}
	\label{eq5}
	\phi (x) = \pm \int\limits_x^\infty {\sqrt { - \frac{2}{r}\frac{d^2r}{dx^2}}
	}
\end{equation}
under supposition of the  integral convergence.

Owing to (\ref{eq3}) the potential $V(x)\equiv V(\phi(x))$ can be expressed by means of $r(x)$:
\begin{equation}
	\label{eq5a}
	V(x) = \frac{1}{r^2} - \frac{A}{r^2}\left( {3\left( r'\right)^2 + rr''} \right) + 2\frac{x - 3m}{r^3}\frac{dr}{dx} .
\end{equation}
Equations (\ref{eq4},\ref{eq5},\ref{eq5a}) represent a general solution in an implicit form by means of  arbitrary $r(x)$ satisfying (\ref{asy_flatness}, \ref{asy_flatness+}). Thus,
we  use the ``inverse'' method  \cite{Bronnikov, BronShikin, BronnikovFabris, Nikonov, Azreg2009, Cadoni2015} to generate families of special solutions: instead of looking for functions $A(x),r(x),\phi(x)$ for some given
potential $V(\phi )$, we may fix any of these functions, for example $r(x)$,
and look for $A(x),\phi (x)$ and $V(\phi )$. This problem is solved by
quadratures according to (\ref{eq4},\ref{eq5},\ref{eq5a}); $V(\phi )$ is defined
parametrically from $V(x)$ using $\phi(x)$.

In case of $r(x)\equiv x$ we have the Schwarzschild metric, the scalar field $\phi\equiv 0$ and $V\equiv 0$.

Further we use  asymptotic relations (see \cite{Azreg2009}) of the solutions depending on the sign of $x_0-3m$. Let the conditions (\ref{asy_flatness}, \ref{asy_flatness+}) for $r(x)$ be satisfied. Then \cite{Azreg2009}

(a)
 if $x_0>3m$, then $A(x)>0, x>x_0$, $A(x)$ and the functions (\ref{eq4}, \ref{eq5}, \ref{eq5a}) have asymptotics as $x\to x_0$:
 \begin{equation}
 \label{Azreg1}
 A(x)\sim \frac{2(x_0-3m)}{3r'(x_0)\, r(x)}, \quad  V(x)\sim \frac{(x_0-3m)\,r''(x_0)}{3r'(x_0)\, r^2(x)};
 \end{equation}
(b) if $x_0<3m$, then there exist a point $x_h>x_0$ (the horizon) such that $A(x_h)=0$ and $A(x)>0$ for $x>x_h$.
\newline

 The case (a) deals with NS. In order to prove (a), one can estimate the Kretschmann scalar near $x_0$
 \begin{equation}
 \label{Kretschmann0}
 R_{\alpha\beta\gamma\delta}R^{\alpha\beta\gamma\delta}\sim
 \frac{16 (x_0-3m)^2}{3[r'(x_0)]^4(x-x_0)^6}.
 \end{equation}
Also, it is easy to show by considering light-like radial geodesics that the time needed for signals from the center $x=x_0$ to reach a remote observer is finite.

In case of $x_0<3m$, it follows from equation (\ref{eq4}) $A(x)\to -\infty$ as $x\to x_0+0$ whence statement (b) follows; and $r(x) > 0, \forall x\ge 0$. There is no any singularities of $\phi, r(x), V(x)$ and the Kretschmann scalar at $x=x_h$. The Schwarzschild-like singularity at $x=x_h$ can be removed by a coordinate transformation, e.g., $(t,x)\to (T,X): T=t+\int dx A^{-1} (1-A)^{1/2},\,\,X=t+\int dx A^{-1} (1-A)^{-1/2}$. In the new coordinates the 2-dimensional surface $x=x_h$ is light-like; therefore, this is indeed the regular horizon.

\section{Test body circular orbits}
\label{orbits}
\begin{figure}[h]
	\includegraphics[width=80mm]{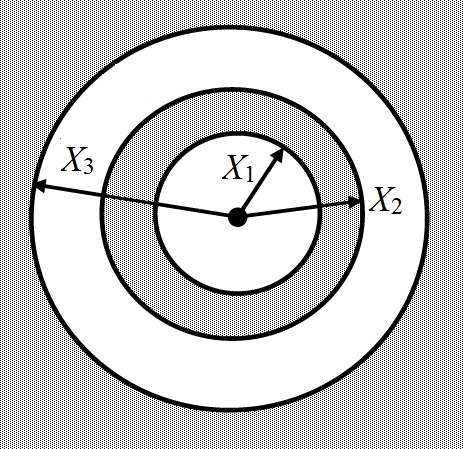}
	\vskip-3mm\caption{Shaded area: regions of stable orbits. White area: no stable orbits. }\label{proba-0}
\end{figure}

In this Section we present general relations to be used below in numerical estimates. We are interested in time-like geodesics in the static region, where $A>0,\, r>0$. We shall concentrate on the case shown in Fig.\ref{proba-0}, and further we use  notations according to this figure. Here $X_1,X_2,X_3$ represent limiting radii of different regions of circular orbits. Namely, the radial coordinates $x: X_3<x<\infty$ correspond to  the outer region of circular orbits, and  $x: X_1<x<X_2$  correspond to  the inner region  (if it exists), which is separated off  the outer region by a prohibited area $X_2<x<X_3$, where either there is no circular orbits or they are unstable.

In case of spherically symmetric space-time metrics (\ref{eq2}) the integrals for the
test body geodesic motion in the equatorial plane are as follows
\[
A(x)\left( {\frac{dt}{d\tau }} \right) = p_t ,\quad r^2(x)\left(
{\frac{d\varphi }{d\tau }} \right) = L,\quad \theta = \pi / 2,
\]
$\tau$ -- is the canonical parameter on the time-like geodesics. These
equations together with the normalization integral yield the equation for
the radial variable
\begin{equation}
\label{eq10}
\left( {\frac{dx}{d\tau }} \right)^2 = p_t^2 - U_{eff} (x,L) ,
\end{equation}
where $U_{eff}(x,L) = L^2 U_1(x) + U_2(x)$,
\begin{equation}
\label{eq11}
U_1 (x) = \frac{A(x)}{r^2(x)},
\quad  U_2 (x) = A(x).
\end{equation}
Note that using a more general metric representation (\ref{eq2.0}) leads to the same equation after the corresponding change of the parameter $\tau$.
Equation (\ref{eq10}) formally describes a one-dimensional  particle motion in a
field with effective potential $U_{eff}/2$.
To study the circular orbits that correspond to the extrema of $U_{eff}$ we shall need the function
\[
  F(x)=-{U}'_2/{U}'_1\,  .
\]
 Making use of (\ref{eq4}) we obtain
 \[
   F(x)=  r^2(x)\left[\frac{2r^3(x)r'(x)}{x - 3m}\int\limits_x^\infty \frac{x' - 3m}{r^4(x')}dx'-1\right]=
 \]
  \begin{equation}
  \label{extrema}
= r^2(x)\left[\frac{r(x)r'(x) A(x)}{x - 3m}-1\right].
    \end{equation}
  Further we use the relations for the derivatives $F',\,F''$  obtained using (\ref{eq4}):
  \begin{equation}
  \label{x_r}
H(x)= F'(x)=\frac{f'(x)A(x)}{2r^2(x)}-4r(x)r'(x)\, ,
  \end{equation}
  \begin{equation}
  \label{X_r_J}
 F''(x)= \frac{f''(x)A(x)}{2r^2(x)}+\frac{r(x)r'(x)}{x-3m}-6r(x)r''(x)-14(r'(x))^2
  \end{equation}
where $f(x)\equiv 2 r^5(x)r'(x)/(x-3m)$ and we assume $r\in C^{(3)}$.

At the points of extrema of $U_{eff}$ we have  $U'_{eff}=0$ yieling
 \begin{equation}
 \label{eq12}
 L^2 = F(x).
 \end{equation}
 If some  $x=X$ is a root of equation (\ref{eq12})
 and   $F(x)$ increases near $X$, then this is a point of a minimum of $U_{eff}$ corresponding to radius $X$ of a stable
 circular orbit with angular momentum $L$; if $F(x)$ decreases, this is the radius of an unstable circular orbit (limiting cycle for the test body trajectories).

 Near the center $x\to x_0+0$ using $r(x)= r'(x_0)(x-x_0)+O(x-x_0)$ we have an asymptotic relation
 \begin{equation}
 \label{Fx_near_center}
 F(x)\sim -\frac{1}{3} r^2(x) .
 \end{equation}
   For $x\to \infty$  using (\ref{asy_flatness}) we have
    \begin{equation}
    \label{Fx_infty}
    F(x)\sim m\,x,\quad H(x)\sim m .
    \end{equation}
Owing to (\ref{Fx_infty}), there is a (sufficiently large) $X_3$ such that for $x\in (X_3, \infty)$ function $F(x)$ is monotonically increasing and any $x$ of this interval is a radius of a stable circular orbit with an appropriate angular momentum $L$.

If $x_0>2m$, then the graph of  $F(x)$ crosses the abscissa axis and tends to infinity as $x\to \infty$; so for any $L$ there is at least one root of (\ref{eq12}).  In particular, there is always a minimum of  $U_{eff}$ for $L=0$, which defines a position of a stationary particle  hanging at rest over the singularity.

In case of  $x_0<3m$ using (\ref{eq4})  we see that
$A(x)>0$ for $x\ge 3m$, including $A(3m)>0$. This means that the horizon $x_h<3m$.  It follows from (\ref{extrema}) that $F(x)\to \infty, \,x\to 3m+0$. For $x\in (x_h,3m)$ we have $F(x)<0$ and there is no roots of  (\ref{eq12}) on this interval. Taking into account the behavior of $F(x)$ as $x\to \infty$, we see that there exists at least one minimum of  $F(x)$
on $(3m,\infty)$.

This can be summarized as follows.

(i) {\it If $x_0<3m$ (the BH case), then there is no circular orbits in the region $x\in (x_h,3m)$. If  for some fixed $L$ we have a maximum of  $U_{eff}$ defining a radius $X_{unst}$ of an unstable circular orbit, then we have a stable circular orbit with radius $X_{st}<X_{unst}$ with the same $L$}, and vice versa, for a stable orbit there is at least one unstable counterpart.

The statement (i) is a simple consequence of the fact that in this case $F(x)$ is monotonically decreasing in the neighborhood of $x=3m$ and monotonically increasing for large $x$.

For example, in case of the Schwarzschild metric  we have $F(x)=x^2m/(x-3m)$; this function has a minimum $F_{min}=12m^2$  for $x=6m$, so there is no solutions of (\ref{eq12}) and  minima of $U_{eff}$ for $L^2<12m^2$.

(ii) {\it For $x_0>2m$ (the NS case) for any $L$ there is always a root $x=X_{st}$ of equation (\ref{eq12}), which corresponds to a minimum of  $U_{eff}$ and defines a radius of a stable circular orbit}.

 In both cases (i) and (ii) there is a solution of (\ref{eq12}) for sufficiently large $L$ and large $x$.

Previous considerations do not say anything about the existence of discontinuous regions of circular orbits.
Let us consider when additional minima of  $U_{eff}$ can appear. This is related to the existence of local maxima of $F(x)$.

(iii) {\it  Let $X_m$ be a point of local maximum of $F(x)$, so that $F(x)\ge 0$ is monotonically increasing for $x\in (X_1, X_m)$ and monotonically decreasing for $x<X_1$ and $x\in (X_m, X'_3)$ for some $X'_3>X_m$. Then there exist  regions of SCO with radii  $x\in (X_1,X_2)$, $X_2=X_m$, and  $x\in (X_3, \infty), \, X_3\ge X'_3$, these regions being separated by the domain with radii $ (X_2,X_3)$,  where SCO do not exist}.

Of course, the point  $X_m$ of (iii) may not  exist. Sufficient conditions for (iii) can be derived in a standard way by investigation of $F(x)$ and its derivatives. For some $X=X_m$ of (iii) to exist,  it is sufficient that
\begin{equation}
\label{cond_1}
F(X)>0\quad \to \quad \frac{r(X)r'(X) A(X)}{X - 3m}>1\,,
\end{equation}
where $A$ is given by (\ref{eq4}),
\begin{equation}
\label{cond_2} F'(X)=H(X)=0,
\end{equation}
where \[
H= \frac{r^2A }{X-3m}\left[ 5(r')^2+r\,r''-\frac{r\,r'}{X-3m}\right]-4rr'\,, \]
(here and in the next equation $r,A,f$ are functions of $X$) and
\begin{equation}
\label{cond_3}
F''(X)<0\,,
\end{equation}
where on account of (\ref{cond_2}) we can put $F''(X)=J(X)$,
\[
J(X)\equiv 4rr'\frac{f''}{f'}+\frac{rr'}{X-3m}-6rr''-14(r')^2
\]
and we have assumed that $r(x)\in C^{(3)}$.

The sufficient conditions (\ref{cond_1}, \ref{cond_2}, \ref{cond_3}) do not define exact boundaries of SCO regions. The limiting radii of such a region are related to a change of the sign of $H(x)=F'(x)$  or to violation of inequality (\ref{cond_1}). Let $(X_1,X_2)$ be a connected SCO region, i.e.  any  $x\in (X_1, X_2)$ is a solution of (\ref{eq12}) for some $L$ such that $x$ is a minimum of $U_{eff}$ and SCO do not exist  in a neighborhood of the region either for $x<X_1$ or for $x>X_2$.

There are two possible types of conditions that will be used  to look for the limiting radii $X_1,X_2$. Here $(X_1, X_2)$ is the interval of monotony of $F(x)$.

Type I:  limiting radii of SCO region are defined by  roots $X_r, \,r=1,2$ of eq. (\ref{cond_2}) satisfying (\ref{cond_1}); $X_1$ is a point of minimum, $X_2$ is a point of maximum of $F(x)$.

Type II:  limiting radius $X_1$ is defined by a root of $F(x)$. This case  corresponds to a minimum of $U_{eff}$ with $L=0$ (minimum of $A(x)$), which describes a particle at rest.\\

Now suppose that $r\equiv r(x,p)$ depends on parameter $p$. Variation of $p$ can change
the conditions listed in (iii) at some bifurcation point $p=p_0$. There are two main types of such bifurcations.

Bifurcation I: the local maximum $F_{max}$ of $F(X)$ crosses zero as $p$ changes. The condition for such a bifurcation to occur is
\begin{equation}
\label{bifur_I}
H(X)=0, \quad F(X)=0, \quad J(X)<0.
\end{equation}
The first two equalities of (\ref{bifur_I}) can be used to find the bifurcation point $p_0$ and $X$-value.

\begin{figure}[h]
	\includegraphics[width=80mm]{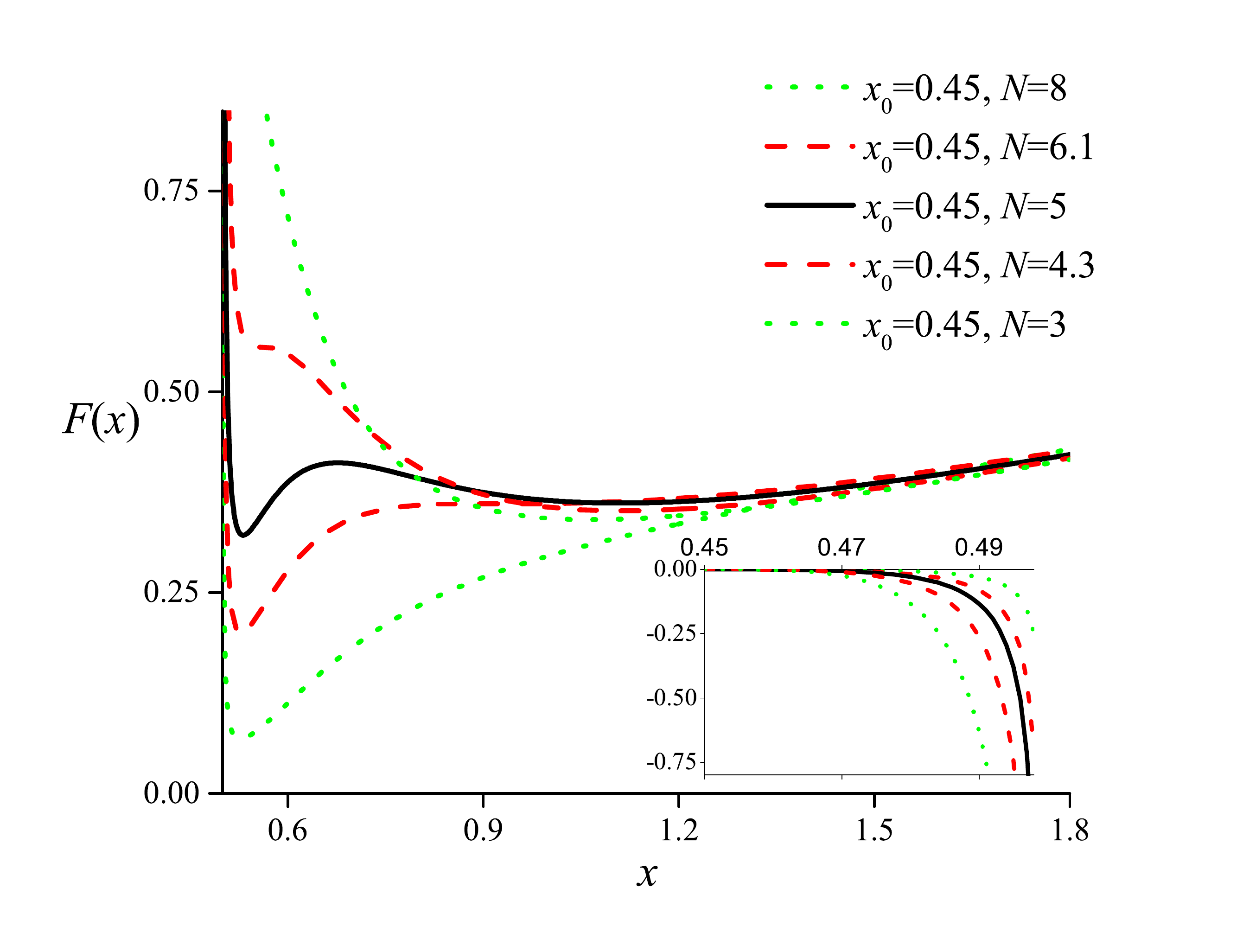}
	\vskip-3mm\caption{The BH case, examples of $F(x)$ with different $N, x_0=0.45$. The main figure shows $F(x)$ for $x>3m$ and smaller graphs in the right lower corner show this function for $x_0<x<3m$, where there is no solutions of (\ref{eq12}). Dashed lines correspond to bifurcation II points c($N_1$) and b($N_2$) in the previous figure, when the maximum of $F(x)$ appears/disappears; solid line -- to $N\in (N_1,N_2)$ when there exists the local maximum, dotted lines -- to $N<N_1$ and $N>N_2$.}\label{exa_F_bh}
\end{figure}

\begin{figure}[h]
	\includegraphics[width=80mm]{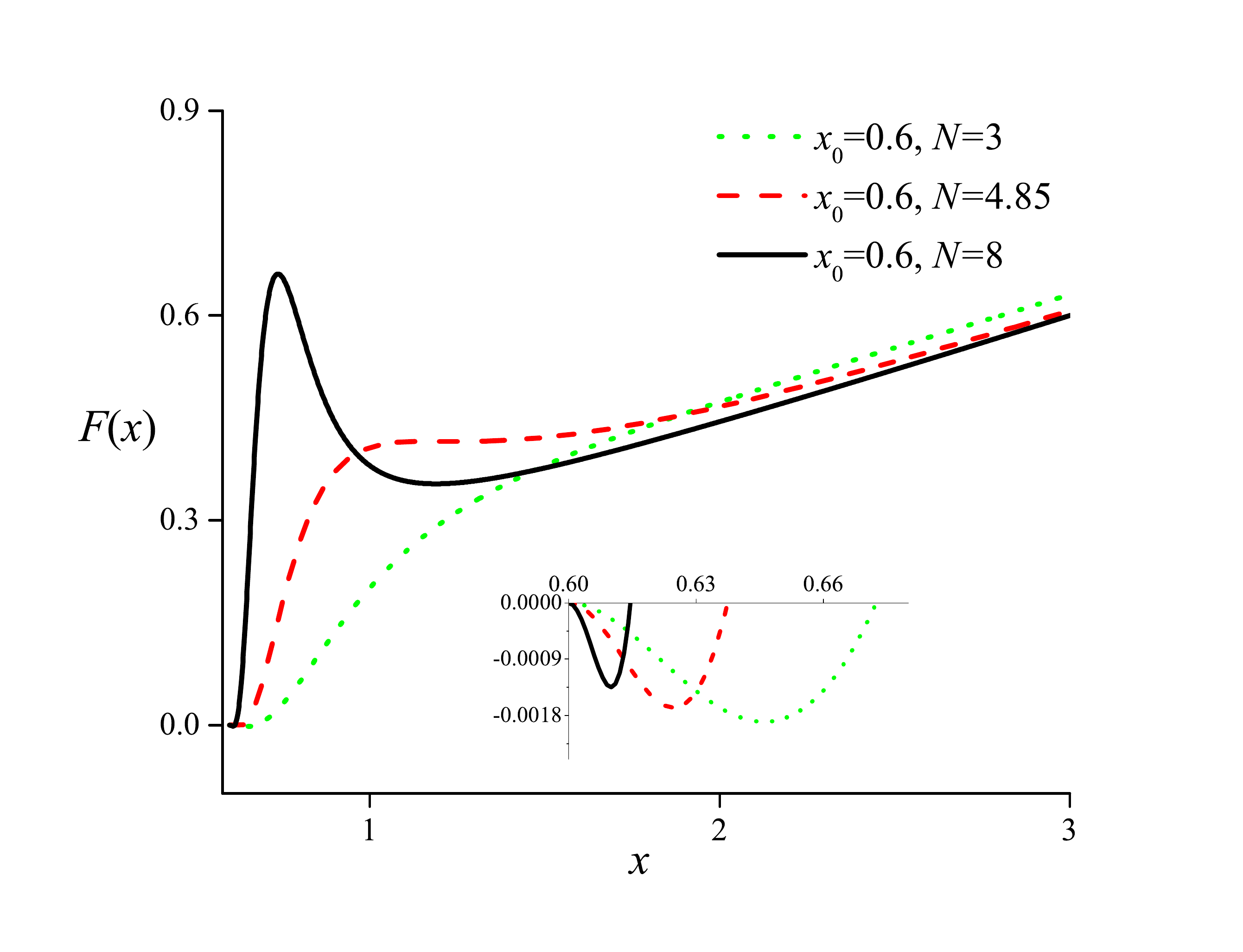}
	\vskip-3mm\caption{The NS case.
	 examples of $F(x)$ with different $N, x_0=0.6$; the smaller graphs in the right lower corner show this function in the neighborhood of $x_0$. Dashed line corresponds to the bifurcation II point "b" in the previous figure ($N_1=4.845$),  when the maximum of $F(x)$ appears/disappears; solid -- to some $N>N_1$ when there exists the local maximum, dotted -- for $N<N_1$ (no  maxima).}\label{exa_F_ns}
\end{figure}

Bifurcation II: the inequality (\ref{cond_1}) remains valid, but the maximum disappears by violating  (\ref{cond_3}), i.e. (necessary condition)
\begin{equation}
\label{bifur_II}
H(X)=0, \quad J(X)=0 , \quad F(X)>0.
\end{equation}
Examples of this bifurcation when maximum of $F(x)$ appears/disappears can be seen in Figs. \ref{exa_F_bh}, \ref{exa_F_ns}.

In a degenerate case one can have $H(X)=0, \, J(X)=0, \, F(X)=0$.

\section{Family of special solutions}
\label{sec:setup}
Looking at conditions (\ref{cond_1}--\ref{cond_3}), we see that there is a functional freedom to fulfill the conditions (iii) by some choice of the arbitrary function $r(x)$. However, it is important to have a simple example on this issue that can be investigated in detail. Having this in mind, we  consider a family of special solutions with
\begin{equation}
	\label{eq6}
	r(x) = x\left[ {1 - \left( {\frac{x_0 }{x}} \right)^N} \right],\quad N > 2.
\end{equation}
In case of $N=2$ we have the example studied in \cite{Azreg2009} (see equation (57) of this paper). The solutions are defined for all $x>x_0$. We shall see that this family is sufficiently wide to include situations with discontinuous SCO distributions.  Note that for $x_0=0$ we have the Schwarzschild solution. The relations (\ref{eq4},\,\ref{eq5},\, \ref{eq5a}) depend on $x_0$ continuously, and this parameter   can be considered as an indicator of the scalar field effects.

Evidently, for $x > x_0 $ we have $r > 0$, $d^2r / dx^2 < 0$ and  the
conditions (\ref{asy_flatness}, \ref{asy_flatness+})  are fulfilled.
Integration in equation  (\ref{eq4})  for $x > x_0 $  yields $A(x)$ in terms of the hypergeometric function
\begin{equation}\label{eq7}
	A(x) = \left[ {1 - \left( {\frac{x_0 }{x}} \right)^N}
	 \right]^2 G(x,x_0,N),
\end{equation}
where
\[  G(x,x_0,N)\equiv
	F\left[ {4,\frac{2}{N},1 + \frac{2}{N},\left( {\frac{x_0 }{x}} \right)^N}	\right] -\]\[
		- \frac{2m}{x}F\left[ {4,\frac{3}{N},1 + \frac{3}{N},\left(
			{\frac{x_0 }{x}} \right)^N} \right] .
\]
In view of (\ref{eq5}) we have
\begin{equation}
	\label{eq8}
	\phi (x) = \pm \sqrt {\frac{8(N - 1)}{N}} \arcsin \left[ {\left( {\frac{x_0
			}{x}} \right)^{N / 2}} \right].
\end{equation}
Formulas (\ref{eq5a},\,\ref{eq7},\,\ref{eq8}) define potential $V(x)$ parametrically for $|\phi|<(\pi/2)\sqrt{8(N-1)/N}$ and correspondingly $x>x_0$. The potential $V(x)$ is explicitly derived using the functions
$A(x),r(x)$ in (\ref{eq5}\,,\ref{eq5a}), then we get $V(\phi )$   by a substitution of
\begin{equation}
\label{x_of_phi}
x = x_0 \left\{ {\sin \left( {\left| \phi \right|\sqrt {\frac{N}{8(N - 1)}}
	} \right)} \right\}^{ - 2 / N}.
\end{equation}
into $V(x)$.

Thus, we get a spherically-symmetric solutions of Einstein's equations  with the scalar field in case of this potential; the solutions are given by (\ref{eq6},\ref{eq7},\ref{eq8}).

\begin{figure}[h]
	\includegraphics[width=80mm]{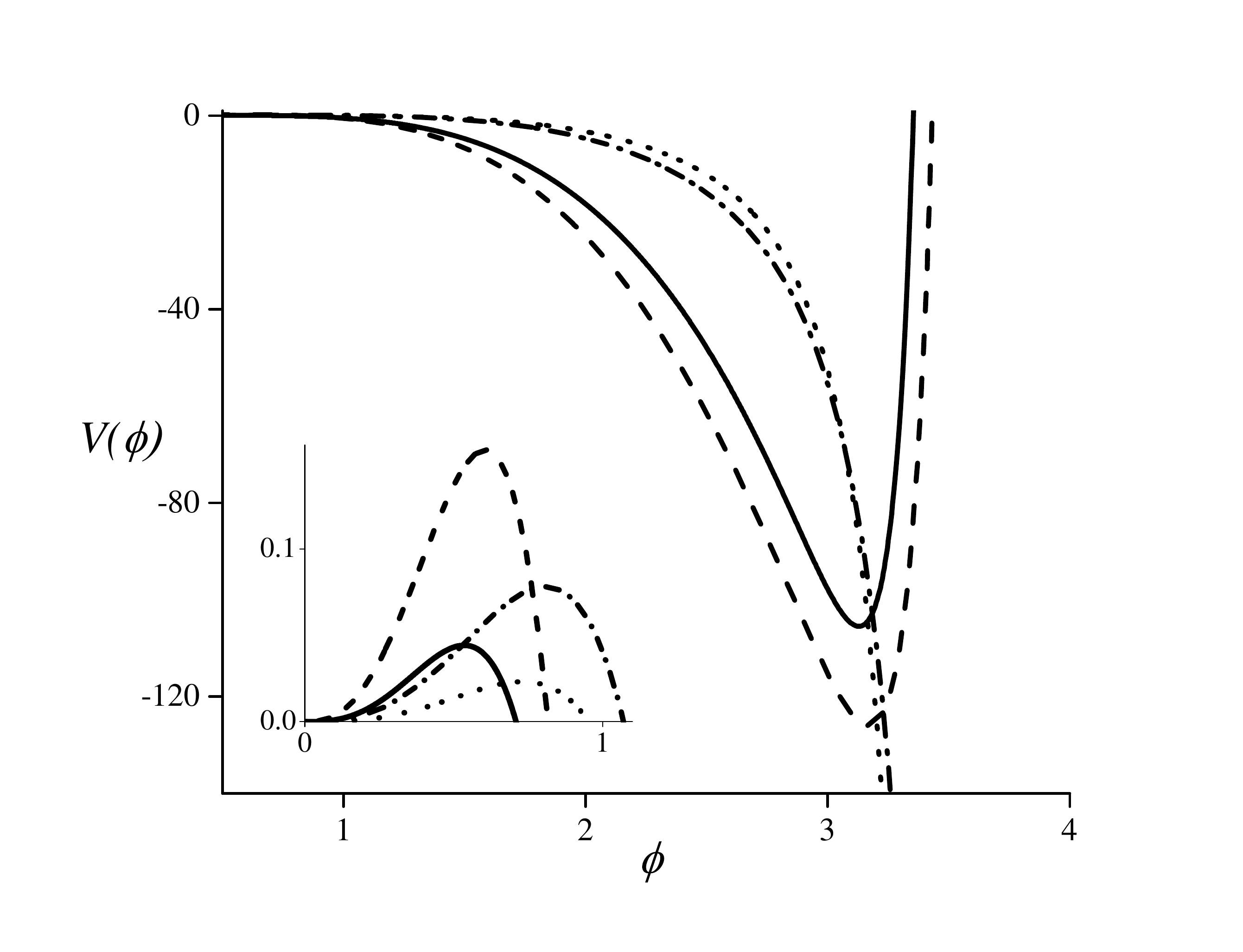}
	\includegraphics[width=80mm]{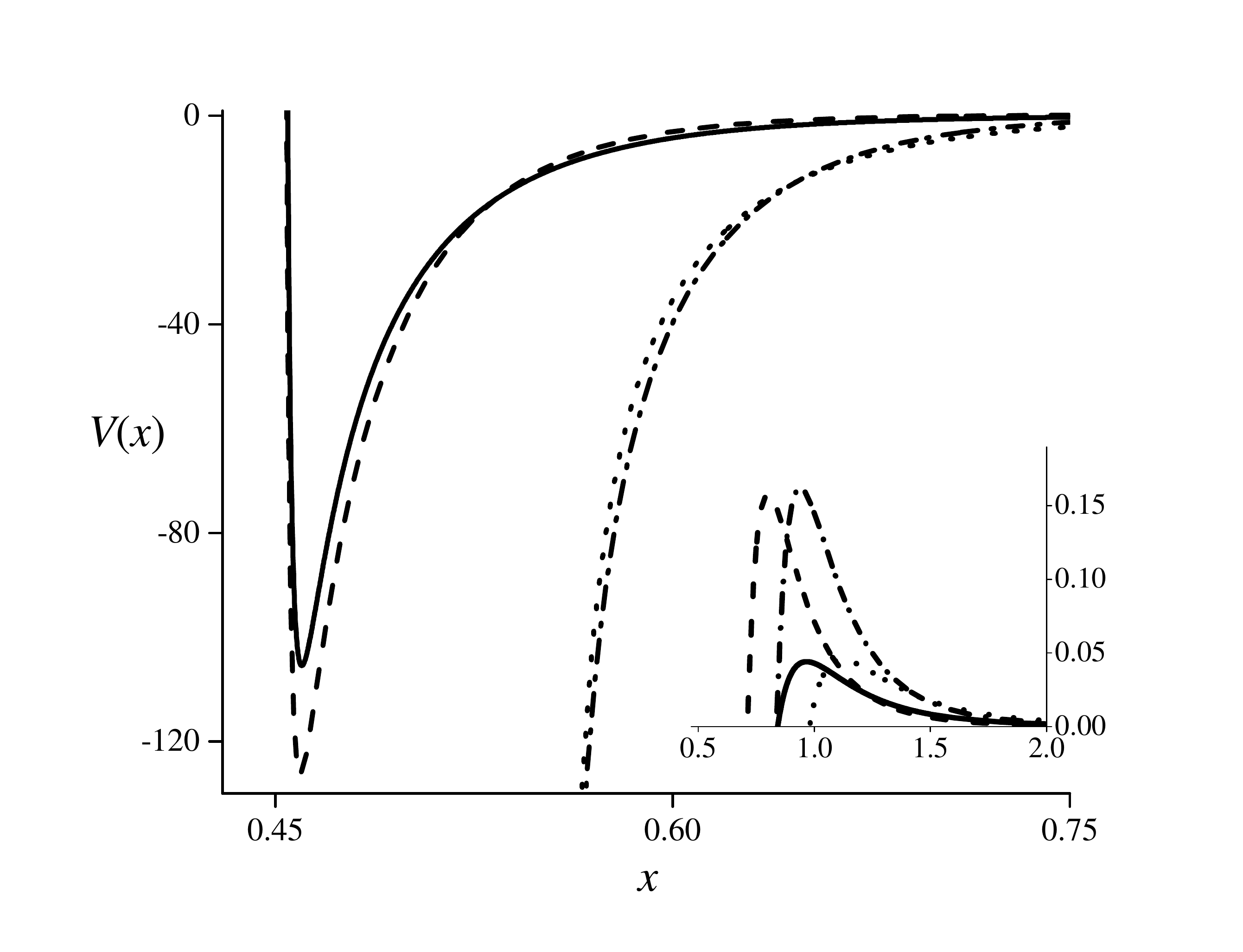}
	\vskip-3mm \caption{Upper panel: scalar field potentials in cases of BH ($x_0=0.45:\, N=4$ -- solid, $N=5$ -- dashed)  and NS  ($x_0=0.55:\,N=4$ -- dotted, $N=5$ -- dash-dot).  Smaller panel shows details of the graphs near the origin $\phi=0$. which cannot be seen on the scale of the large panel. We note that the potential graph of the intermediate case $x_0=0.5$ (not shown here)  is completely similar to the case of $x_0>0.5$. \\
		Lower panel: the same potentials $V(x)\equiv V(\phi(x))$ as functions of $x$. Smaller panel at the right corner shows details of the graphs for large $x$.}\label{SF_Potentials}
\end{figure}

The scalar field potentials are qualitatively different is cases of BH and NS.
As $|\phi| \to \pi \sqrt {2(1 - 1 / N)}$, in case of NS we have $V(\phi ) \to - \infty $ and in case of BH $V(\phi ) \to \infty$. In the latter case the  graph of the potential is similar to the "Mexican hat" (Fig. \ref{SF_Potentials}).
In both cases the potentials $V(\phi)$ are negative in some  region (see examples in Fig. \ref{SF_Potentials}), so our solutions do not contradict to the no-scalar-hair theorems \cite{Bekenstein, Bronnikov}).
Asymptotic relations for the solutions at spatial infinity and near the center  can be derived either directly from the formulas of Section \ref{basic}, or from (\ref{eq8}, \ref{x_of_phi}); they are listed in Appendix \ref{asymp-massless}. The behavior at spatial infinity (large $x$) corresponds  to the behavior of $V(\phi)$ near the origin $\phi=0$. Correspondingly, we get the scalar field self-interaction potential near $\phi=0$; it is asymptotically the same for the NS and BH cases (see Fig. \ref{SF_Potentials}):
\begin{equation}
\label{V_of_phi_0}
V(\phi)\sim \frac{(N-2)N^{2(1+1/N)}}{(N-1)^{2/N}(N+2)\,x_0^2}
{\left(\frac{|\phi|}{2\sqrt{2}}\right)}^{2(1+2/N)}.
\end{equation}
Thus, we are dealing with massles scalar fields.

The Kretschmann invariant diverges at the center; in accordance with (a) and (b) of Section \ref{basic}, for $x_0 > 3m$ there is the NS at the center $x=x_0$; and for $x_0 < 3m$ we have BH with a horizon at some $x=x_{h}>x_0$.

\section{SCO distribution in case of (\ref{eq6})}
\label{numerical}
For numerical estimates it is convenient to choose further the units of length so as to have $C=6m=1$. Using (\ref{cond_1}, \ref{cond_2},\ref{cond_3}) one can look for parameters $x_0>0,\,N>1$ leading to discontinuities in the SCO distribution.
 After determination of roots $X_r$ of $H(x)$ and  $F(x)$ for fixed $x_0,N$, we determine limiting values of angular momentum $L_{lim}$ according to (\ref{eq12}) when  minima of $U_{eff}$ begin to appear/disappear. In this problem both type I (roots of $H(x)$) and type II (roots of $F(x)$) of limiting SCO can be present.

\begin{figure}[h]
 	\includegraphics[width=80mm]{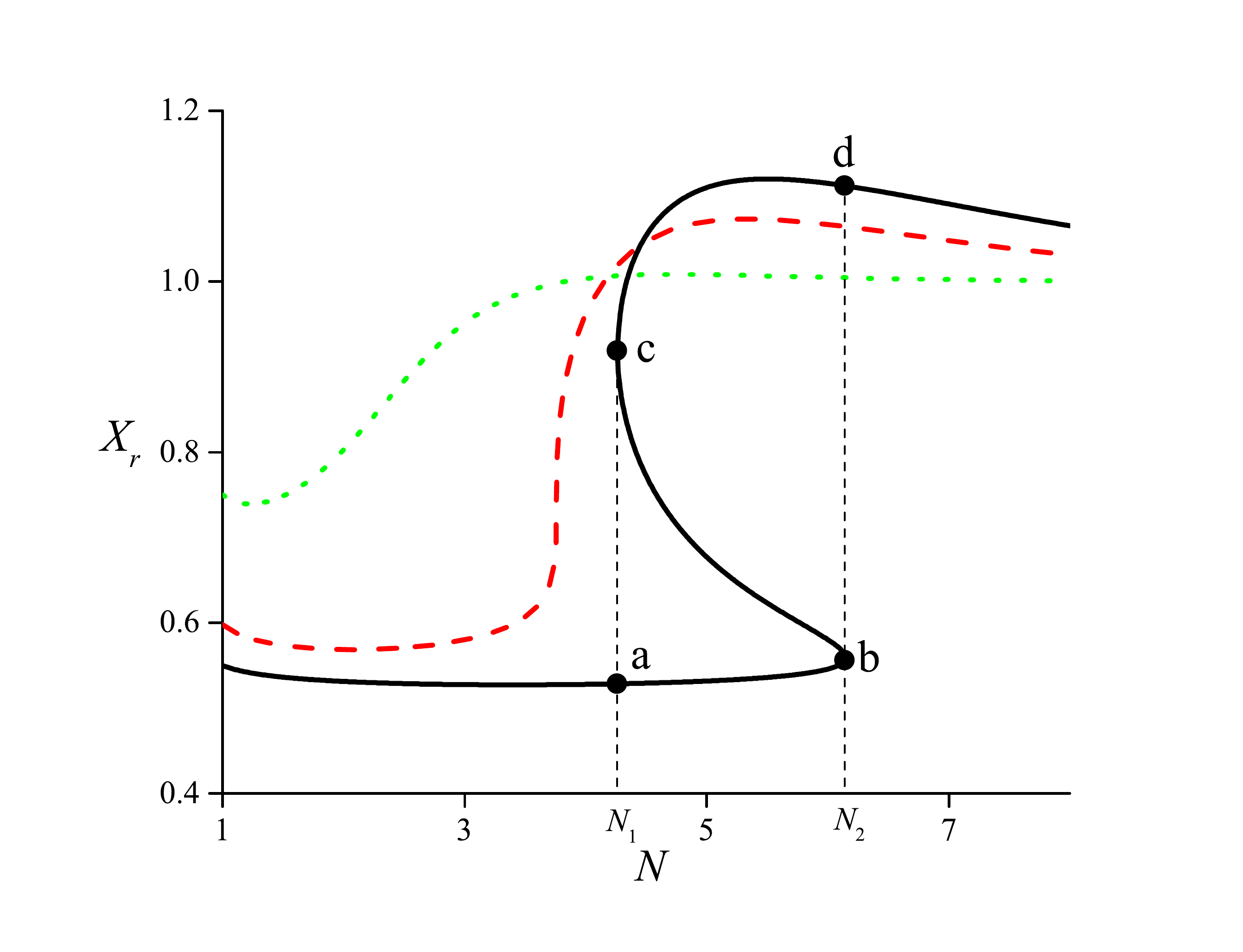}
 	\vskip-3mm\caption{The BH case, limiting SCO radii $X_r(N)$ (type I) for three values of $x_0$. The  dotted curve shows the single-valued dependence $X_1(N)$ for  $x_0=0.25$. The case of $x_0=0.40$ (dashed) is critical: for larger $x_0$ we  have three roots of equation (\ref{cond_2}). Correspondingly, the case of $x_0=0.45$ (solid) represents  three-valued function $X_r(N)$ between $N_1= 4.3$ and $N_2= 6.1$: the branch between points "a" and "b" represents $X_{1}(N)$, between "b" and "c" -- $X_{2}(N)$ and between "c" and "d" -- $X_{3}(N)$ (notations according to Fig. \ref{proba-0}); there is no SCO radii between  $X_{2}(N)$ and  $X_{3}(N)$. The latter branch extends to  infinity for $N>N_2$. The points c($N_1$) and b($N_2$) represent the bifurcation II points.   }\label{xk_of_N_bh}
 \end{figure}

\begin{figure}[h]
	\includegraphics[width=80mm]{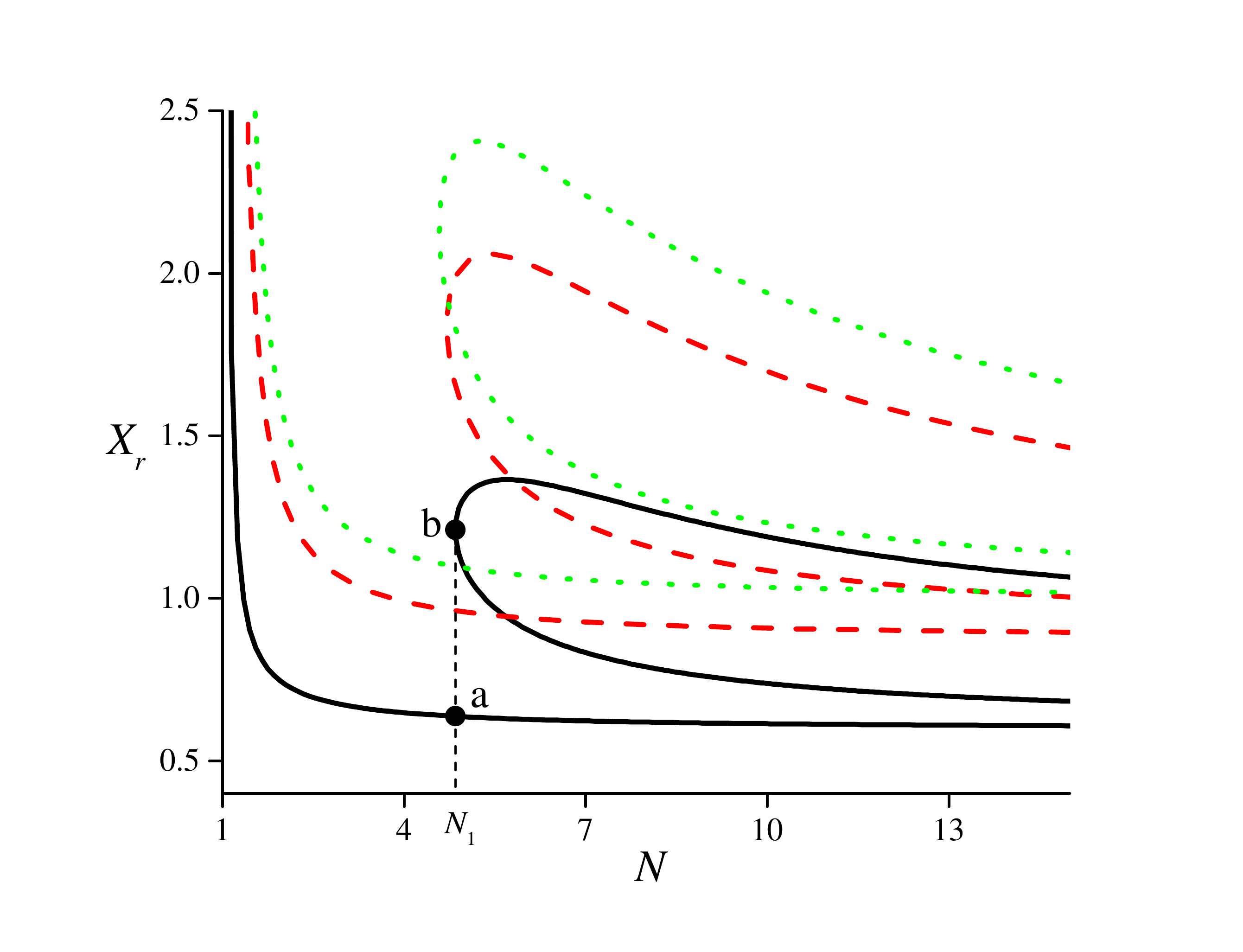}
	\vskip-3mm\caption{The NS case, dependencies $X_r(N),  \,r=1,2,3$ for  $x_0=0.6$ (solid), $x_0=0.88$ (dashed): $x_0=1$  (dotted). The lower curves $X_1(N)$ represent type II  limiting SCO radii. Two upper curves $X_2(N), X_3(N), N>N_1$ represent type I limiting SCO radii; these curves are connected at bifurcation II points. E.g., for $x_0=0.6$, to the right of the line "ab" (with abscissa $N_1=4.845$) we have configurations with discontinuous distribution of SCO; and  there is no SCO radii between  $X_{2}(N)$ and  $X_{3}(N)$.}\label{xk_of_N_ns}
\end{figure}

\begin{figure}[h]
	\includegraphics[width=90mm]{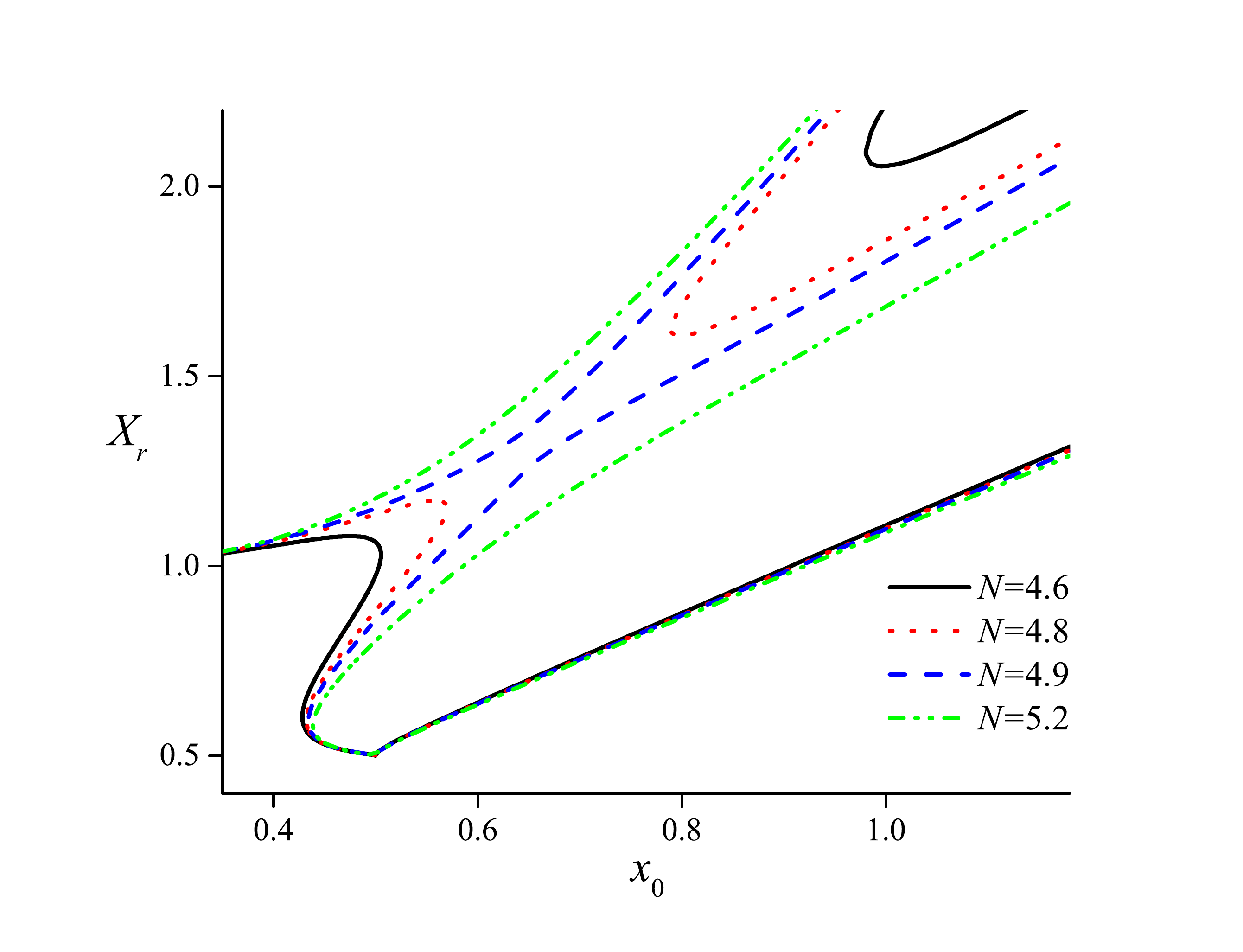}
	\vskip-3mm\caption{$X_r(N)$ as a function of $x_0$ for different $N$ in the neigborhood of  critical  value $N_c=4.87$, when continuous line ($N<N_c$) breaks into non-connected branches. The lower sections of the curves for $x_0>0.5$  (the NS case), which  almost have merged together, represent the type II limiting SCO radii. The other sections of the curves correspond to the type I limiting SCO.}\label{xk_of_x0}
\end{figure}

 Further we present  results of the numerical investigation; they look different for BH ($x_0<0.5$) and NS ($x_0\ge 0.5$). The dependences of boundary radii  $X_r,\,r=1,2,3$ upon $N$ for some fixed values of  $x_0$ are illustrated in Figs. \ref{xk_of_N_bh}, \ref{xk_of_N_ns}.  Fig. \ref{xk_of_x0} represents the dependence  $x_k(x_0)$ for some fixed values of  $N$.

(i) In case of BH $U_{eff}(x_h,L)=0$ and function $U_{eff}(x,L)$ is increasing as a function of $x$ in some neighborhood of $x_h$. The type of monotony for larger $x>x_h$ (and the occurrence of double minima of $U_{eff}$) depends on $L,x_0,N$.  The curves in this case are represented as the roots $X_{r}$ of $H(x)$.


For sufficiently  small  $x_0$ (small scalar field effects) a typical situation is represented by the dotted curve $X_1(N)$ of Fig. \ref{xk_of_N_bh}. Here the dependence $X_1(N)$ single-valued, that is for every $N$ there is the only  root $X(N)$  that defines a lower  boundary of SCO radii and a corresponding value of the limiting angular momentum $L_{lim}=\sqrt{F(X(N))}$.
 In this case there is the only connected region of SCO, different SCO having different angular momenta. This region contains non-relativistic SCO with large $x$ and correspondingly large $L$.

For larger $x_0$ the dependence $X_r(N)$ becomes many-valued between some  $N_1,N_2$. A typical picture  is represented by the solid curve of Fig.\ref{xk_of_N_bh}; "c" and "b" represent bifurcation II points. The lower section of the curve  $x=X_{1}(N)$ is defined for $N<N_2$.
For $N_1<N<N_2$  there are three branches: $X_{1}(N)<X_{2}(N)<X_{3}(N)$ that are connected at points "b" and "c". For any fixed $N\in (N_1,N_2)$ these dependencies define two intervals of the SCO radii: $(X_{1},X_{2})$ , $(X_{3},\infty)$;
  the intermediate region $(X_2,X_3)$   is being prohibited (see Fig. \ref{proba-0}).
For $N>N_2$ there is again only one root $X_{3}(N)$ of $H(x)$.

(ii) In case of NS  there also can be disconnected regions of the SCO. Typical dependences $X_r(N)$ in the NS case are presented  in Fig. \ref{xk_of_N_ns}. We see that for all $x_0>3m$ there exists $N_1$ such that for $N<N_1$ there is the only branch of the type II limiting radii that extends to infinity. For  $N>N_1$ additional two branches $X_{2}$ and $X_{3}$ appear (type I limiting SCO, roots of $H(x)$).  The prohibited region  occupies the space between $X_{2}$ and $X_{3}$ that exist not for all parameters of the family.

The  dependences $X_r(x_0)$ plotted in Fig. \ref{xk_of_x0} represent two  qualitatively different cases.   For large $N$ there  always exists an upper branch $X_3(x_0)$ for all $x_0$. The continuous lower curve always consists of two parts  connected at the  point $x_0=0.5$ with a break. For $N<4.87$  this curve is broken into two parts; the smaller $N$, the farther away is the right upper branch of the curve in the region of large $x_0$.
In the same way as in Fig. \ref{xk_of_N_ns}, the lower curve in Fig. \ref{xk_of_x0} to the right of the break point  also represents type II, not  not type I.
The other curves in this figure are obtained as the roots $X_{r}$ of $H(x)$ (type I).

\section{Some generalizations}
\label{generalizations}
The relations (\ref{cond_1}--\ref{cond_3}) are local conditions on functions $r(x),r'(x),r''(x),r'''(x)$ and $A(x)$; the latter continuously depends on $r(y)$ for $y\in [x,\infty)$. Let
\begin{equation}
\label{eq6gen2}
r(x) = x\left[ 1 -  \varepsilon(x) \right].
\end{equation}
It is easy to see that small variations of  $\varepsilon(x)$ and its derivatives do not change the topological structure of SCO distribution. Therefore, we may construct infinity of examples
leading to a discontinuous structure similar to that described in Fig. \ref{proba-0} using these variations. A simple outcome is, in particular, can be formulated as follows. Let $C=6m$ is fixed,  $\varepsilon\in C^{(3)}$,   $||\varepsilon||=\sup\{|\varepsilon(x)|+x|\varepsilon'(x)|+x^2 |\varepsilon''(x)|+x^3|\varepsilon'''(x)|, \,x\in [x_0,\infty)\} $. {\it  There exists a sufficiently small value   $\varepsilon_0>0$, such that if $||\varepsilon||<\varepsilon_0$, then there are no  discontinuities in the SCO distribution}. This statement follows from inequality $F''(x)>0$ (which contradicts to \ref{cond_3}) in the neighbourhood of minimum of $F(x)$ for small variations near the Schwarzschild metric (see Section \ref{orbits}).

\begin{figure}[h]
	\includegraphics[width=90mm]{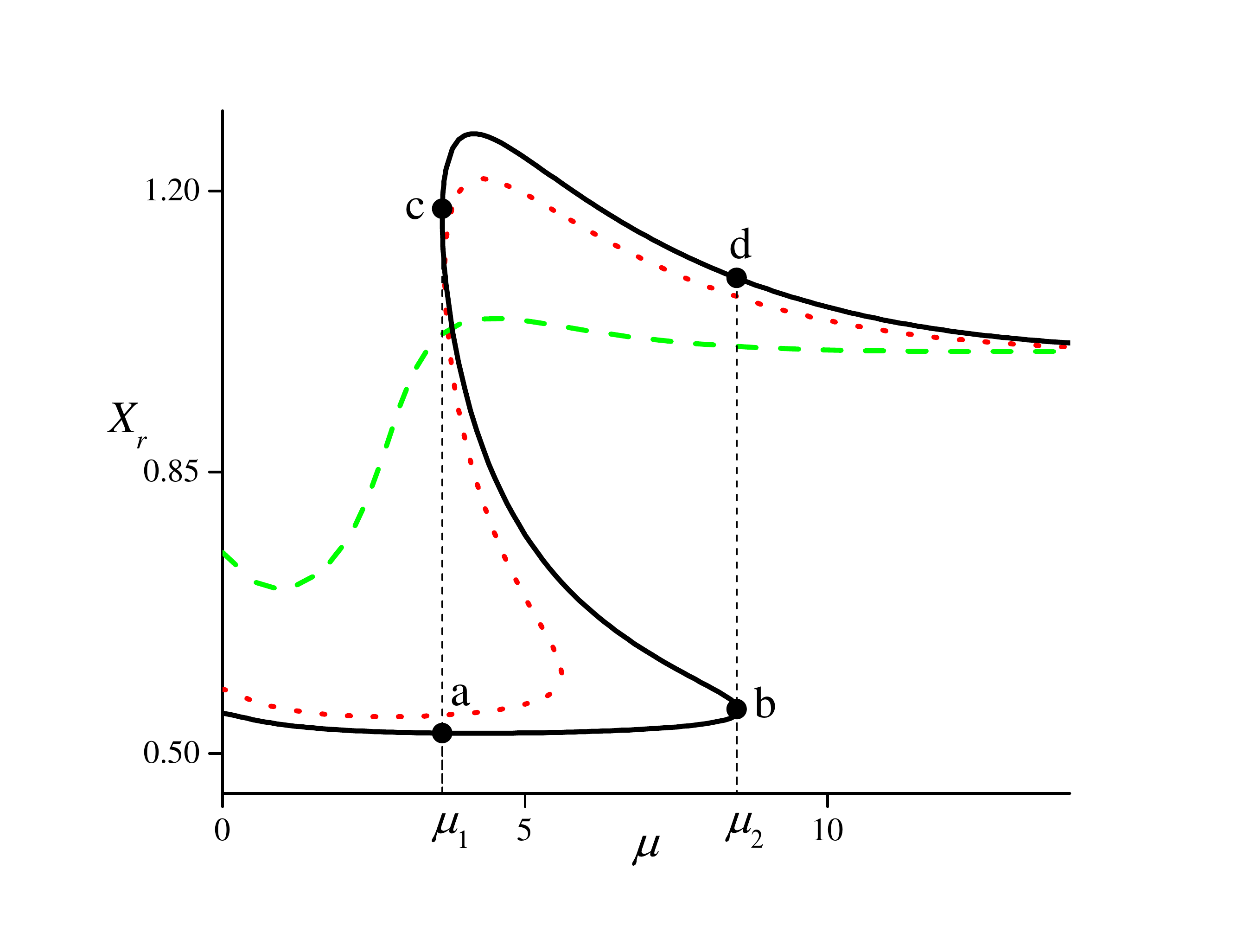}
	\vskip-3mm\caption{Limiting SCO radii $X_r(\mu)$ for three values of $x_0$ in the BH case of the generating function (\ref{eq6gen2}). The dashed curve shows $X_1(\mu)$ for  $x_0=0.25$, which defines radii of the innermost SCO $X_{1}(\mu)$. The case of $x_0=0.45$ (solid) represents a three-valued function between $\mu_1\approx 3.6$ and $\mu_2\approx 8.5$: the branch between points "a" and "b" represents $X_{1}(\mu)$, between "b" and "c" -- $X_{2}(\mu)$ and between "c" and "d" -- $X_{3}(\mu)$ according to Fig. \ref{orbits}.  The case of $x_0=0.42$ (dotted line) is intermediate, it shows how the  curves transform as $x_0$ changes. }\label{xk_of_mu_bh}
\end{figure}

\begin{figure}[h]
	\includegraphics[width=90mm]{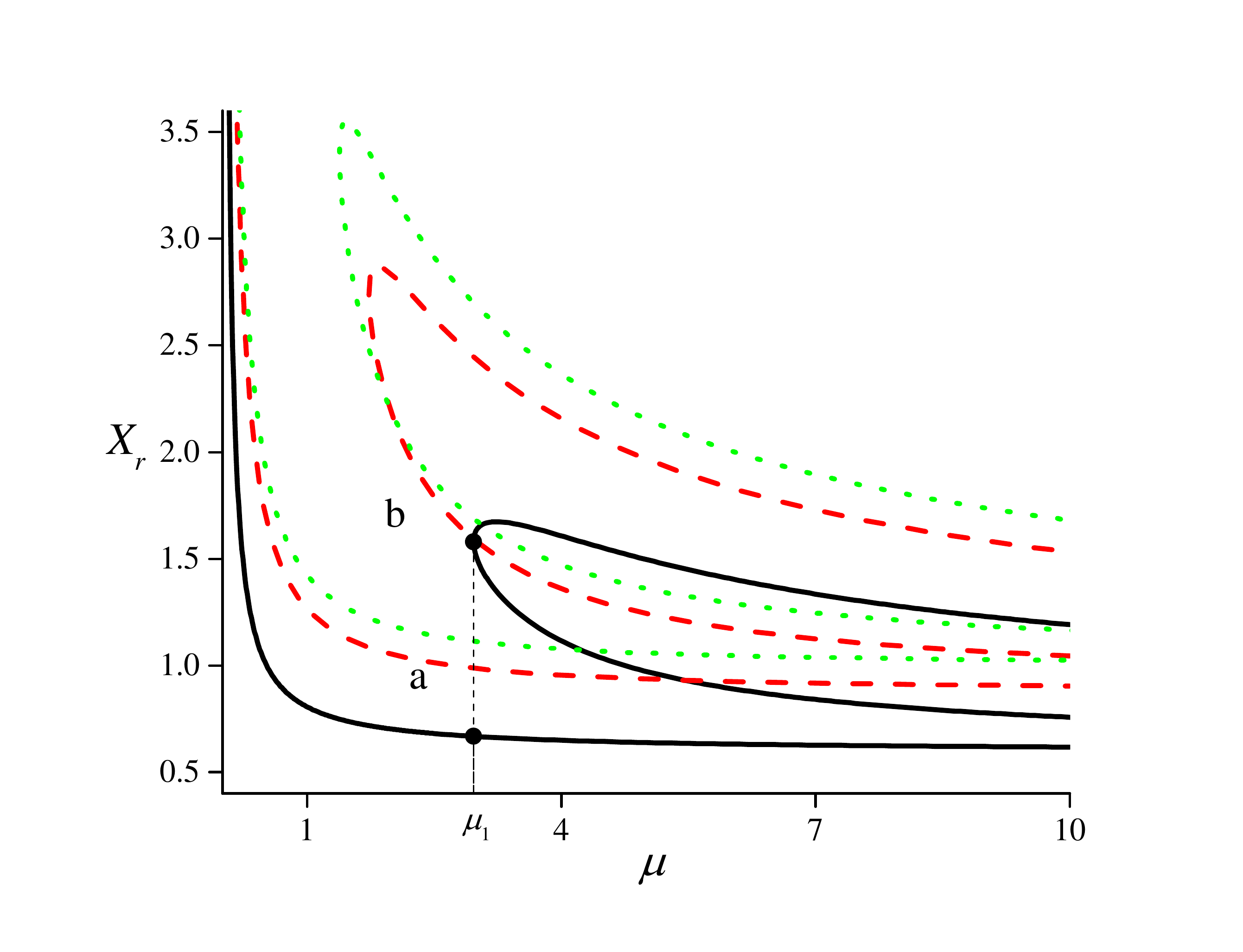}
	\vskip-3mm\caption{The case of NS of the generating function (\ref{eq6gen2}): dependencies $X_r(\mu)$ for  $x_0=0.6$ (solid), $x_0=0.88$ (dashed): $x_0=1$  (dotted). For fixed $x_0$ and $\mu >\mu_1$ (to the right of line "ab") there exists the region of SCO with radii in the interval $(X_{1}(\mu),X_{2}(\mu))$ and $(X_{3}(\mu), \infty$.  The  curves  $X_{2}, X_3$ and $X_{1}(\mu)$ represent correspondingly type I and type II limiting SCO radii; "b" is the bifurcation II point.}\label{xk_of_mu_ns}
\end{figure}

Now we consider a modification of  (\ref{eq6}) assuming
\begin{equation}
\label{eq6gen22}
\varepsilon(x) =   \left( {\frac{x_0 }{x}} \right)^N e^{-\mu (x-x_0) } .
\end{equation}
The main interest to (\ref{eq6gen22}) is due to asymptotic behavior for small $\phi$, which is different from that of  (\ref{V_of_phi_0}) in case of  (\ref{eq6}). This modification generates the scalar field potential with asymptotic behavior
\begin{equation}
\label{V_of_phi}
V(\phi)\sim \frac{\mu^2}{8}\phi^2
\end{equation}
for small $\phi$ (and any fixed $N$), corresponding to the scalaron mass $\mu/2$; this follows from asymptotic relations at spatial infinity (Appendix \ref{asymp-massive}). Apart this property, the graphs and qualitative behavior of $V(\phi)$  both in the BH and NS cases are similar to those of Section \ref{sec:setup}. Here we present  examples of the dependencies $X_r(\mu)$ (Figs. \ref{xk_of_mu_bh},\,\ref{xk_of_mu_ns}). These figures also are  qualitatively similar  to upper panels of Figs. \ref{xk_of_N_bh},\, \ref{xk_of_N_ns}. For a numerical example (Figs. \ref{xk_of_mu_bh},\,\ref{xk_of_mu_ns}) below we have chosen $N =1+m\mu$ in order to compare with some of the results of  \cite{Asanov}.

In the BH case, for sufficiently  small  $x_0$ there is the only connected region of possible SCO and we have a single-valued  dependence $X_r(\mu)$ that defines innermost SCO.
For larger $x_0$ there arise two bifurcation II points such that  the dependence $X_r(\mu)$ becomes many-valued between some  $\mu_1,$ $\mu_2$ (see solid line in Fig. \ref{xk_of_mu_bh}, and the region between perpendiculars "ac" and "bd" to the abscissa axis).
For $\mu>\mu_2$ there is again the only connected region of possible SCO.

In case of NS for fixed $x_0>0$ (whatever small) there exists $\mu_1>0$, such that for $\mu>\mu_1$ the disconnected regions of the SCO arise (see a region to the right of "ab" and solid line in Fig. \ref{xk_of_mu_ns}). Typical dependences $X_r(\mu)$ in the NS case are presented  in Fig. \ref{xk_of_mu_ns}. The lower branch represents the type II limiting radii. The other two branches for $\mu>\mu_1$  represent the type I ones.

The similarity of $X_r(N)$ and $X_r(\mu)$ for (\ref{eq6}) and (\ref{eq6gen22}) correspondingly suggests that occurrence of discontinuous SCO distribution is related to how fast the field $\phi(x)$ and the potential $V(\phi(x))$  decrease as $x$ increases. The rate of the decrease is obviously depends on the rate of decrease of $\varepsilon(x)$, which is defined by parameters $N$ and $\mu$. This property looks different in cases of NS and BH: in the former case, the non-connected SCO distributions exist for all $N$ and/or $\mu$ up to infinite values. In case of BH the corresponding interval of $N$ and/or $\mu$ is limited.

\section{Discussion}
\label{conclusions}
We presented special solutions of the joint system of Einstein equations and massless scalar field equations with various nonzero self-interaction  potentials. The  solutions describe isolated static spherically symmetric configurations with an  asymptotically flat space-time and a positive total mass.
These solutions deal with either NS in the center of the configuration, or with BH. The family includes the Schwarzschild metric as limiting case $x_0=0$; the smaller $x_0$, the closer the solutions to the Schwarzschild case. In the BH case, the scalar field potential $V(\phi)$ is bounded from below; and the potentials are infinitely negative near NS.

The main outcome of this paper is that separated (disconnected) ring-like structures of stable circular orbits can indeed emerge for some family parameters, no matter BH or NS case, massless or massive case. Note that such a feature in case of the massless linear scalar field was first revealed in
 \cite{Chowdhury} dealing with NS at the center. Our considerations show that similar structures can emerge also in BH cases. We remind
that the well-known BH-no-hair theorem \cite{Bekenstein} does not prohibit the BH case, if the  scalar field potential is not positively definite; this is just the case of our solutions.
It is also important to note (see Figs. \ref{xk_of_N_ns}, \ref{xk_of_mu_ns}) that occurrence of NS does not necessarily imply the existence of the discontinuous structures. As to more fine effects due to the form of the scalar field potential, the situation looks rather complicated. There are some considerations concerning the rate of decrease of $V(x)$ (see Section \ref{generalizations}), but they rely upon the concrete form of $r(x)$.

Evidently, our results dealing with a special family solutions do not describe the most general situation with the scalar field. Moreover, our solutions show that for some family parameters the discontinuous distribution of SCO  exists and for the other (in particular, when the parameter $x_0$ is sufficiently small) -- it does not exist.  Nevertheless, the results suggest that the existence of separated annular regions in accretion disks around  relativistic objects can be a fairly common phenomenon that may  indicate some deviations from standard models.

 This effect can be tested by means of   the fluorescent
iron lines (especially, Fe~K$\alpha$) in the X-ray spectra radiated by accretion disks around compact objects  \cite{Guilbert, Lightman, Fabian}. The appearance of the separated ring-like structures in astrophysical objects could be detected by unusual deformations of these lines. Of course, the scalar field effects, if they exist, will be obscured by less exotic ones, e.g., due to accretion disk warps in AGNs, inhomogeneous distributions of matter in the disks, reflection of X-rays from dust tori,  and possible existence of BH companions (see, e.g., \cite{Vasylenko, Fedorova}).  This will complicate the  comparison of effects associated with different kinds of exotic objects \cite{Bambi}.

\acknowledgments
Authors are  thankful to the referees of this paper for helpful suggestions. This work has been supported in part by   State Foundation for Fundamental Research in Ukraine. O.S. is indebted for a partial support from Department of target training of Taras Shevchenko National University of Kyiv under National Academy of Sciences of Ukraine (project 6F).


\appendix
\section{Asymptotic relations: massless case}
\label{asymp-massless}
Asymptotic relations at the spatial infinity are as follows
\begin{equation}
\label{A_at_infty}
A(x)= 1-\frac{2m}{x}+\frac{2(2-N)}{N+2}\left(\frac{x_0}{x}\right)^N \left[1+O\left(\frac{1}{x}\right) \right]\,,
\end{equation}
\begin{equation}
\label{V_at_infty}
V(x)= \frac{N(N-1)(N-2)}{N+2}\frac{x_0^N}{x^{N+2}}\left[1+O\left(\frac{1}{x}\right)  \right] .
\end{equation}
These relations can be used to obtain formula (\ref{V_of_phi_0}).

Using (\ref{eq4}) and (\ref{eq6})  we have in the
neighborhood of $x_0 $ the metric coefficient
\begin{equation}
\label{eq9}
A(x) \sim \frac{2(x_0 -3m)}{3N^2(x-x_0)}
\end{equation}
and the potential
\begin{equation}
\label{V_center}
V(x) \sim -\frac{(N-1)(x_0 - 3m)}{3N^2x_0(x-x_0)^2} .
\end{equation}

\section{Massive scalar field}
\label{asymp-massive}
Here we present asymptotic relations in case of (\ref{eq6gen2}). At spatial infinity we have
\begin{equation*}
A(x)= 1-\frac{2m}{x}-
\end{equation*}
\begin{equation}
-2\varepsilon\left[1-\frac{2m\mu+4 }{\mu x}+O\left(\frac{1}{x^2}\right)\right]+O\left(\varepsilon^2\right),
\end{equation}
where $\varepsilon=\varepsilon(x) $ is given by (\ref{eq6gen22}),
\begin{equation}
\phi(x)=\pm 2\sqrt{2\varepsilon}\left[1-\frac{1}{\mu x}+O\left(\frac{1}{x^2}\right)\right]
+O\left(\varepsilon^{3/2}\right),
\end{equation}
\begin{equation}
V(x)=\mu^2 \varepsilon \left[1+\frac{2N-2m\mu-6}{\mu x}+O\left(\frac{1}{x^2}\right) \right] +O\left(\varepsilon^2\right),
\end{equation}
Behavior near the center:
\begin{equation}
A(x)\sim\frac{2(x_0-3m)}{3(N+\mu x_0)^2(x-x_0)} \,,
\end{equation}
\begin{equation}
V(x)\sim -\frac{(x_0-3m)((\mu x_0)^2+(N-1)(N+2\mu x_0))} {3x_0(x-x_0)^2(N+\mu x_0)^3} .
\end{equation}


%


\begin{thebibliography}{0}%
\makeatletter
\providecommand \@ifxundefined [1]{%
 \@ifx{#1\undefined}
}%
\providecommand \@ifnum [1]{%
 \ifnum #1\expandafter \@firstoftwo
 \else \expandafter \@secondoftwo
 \fi
}%
\providecommand \@ifx [1]{%
 \ifx #1\expandafter \@firstoftwo
 \else \expandafter \@secondoftwo
 \fi
}%
\providecommand \natexlab [1]{#1}%
\providecommand \enquote  [1]{``#1''}%
\providecommand \bibnamefont  [1]{#1}%
\providecommand \bibfnamefont [1]{#1}%
\providecommand \citenamefont [1]{#1}%
\providecommand \href@noop [0]{\@secondoftwo}%
\providecommand \href [0]{\begingroup \@sanitize@url \@href}%
\providecommand \@href[1]{\@@startlink{#1}\@@href}%
\providecommand \@@href[1]{\endgroup#1\@@endlink}%
\providecommand \@sanitize@url [0]{\catcode `\\12\catcode `\$12\catcode
  `\&12\catcode `\#12\catcode `\^12\catcode `\_12\catcode `\%12\relax}%
\providecommand \@@startlink[1]{}%
\providecommand \@@endlink[0]{}%
\providecommand \url  [0]{\begingroup\@sanitize@url \@url }%
\providecommand \@url [1]{\endgroup\@href {#1}{\urlprefix }}%
\providecommand \urlprefix  [0]{URL }%
\providecommand \Eprint [0]{\href }%
\providecommand \doibase [0]{http://dx.doi.org/}%
\providecommand \selectlanguage [0]{\@gobble}%
\providecommand \bibinfo  [0]{\@secondoftwo}%
\providecommand \bibfield  [0]{\@secondoftwo}%
\providecommand \translation [1]{[#1]}%
\providecommand \BibitemOpen [0]{}%
\providecommand \bibitemStop [0]{}%
\providecommand \bibitemNoStop [0]{.\EOS\space}%
\providecommand \EOS [0]{\spacefactor3000\relax}%
\providecommand \BibitemShut  [1]{\csname bibitem#1\endcsname}%
\let\auto@bib@innerbib\@empty
\end{thebibliography}%


\begin{thebibliography}{73}%

\makeatletter

\bibitem[\protect\citeauthoryear{Will}{2014}]{Will}
C.M. Will. The Confrontation between General Relativity and Experiment. Living Rev. Relativ. {\bf 17}, 4 2014 [arXiv:1403.7377]

\bibitem[\protect\citeauthoryear{Berti}{2015}]{Berti} E. Berti,  E. Barausse, V. Cardoso, et al. Testing General Relativity with Present and Future Astrophysical Observations
Classic. Quant. Grav., {\bf 32}, 243001 (2015) 	[arXiv:1501.07274]

\bibitem[\protect\citeauthoryear{Plank}{2013}]{Plank} Plank collaboration. Planck 2013 results. XXII. Constraints on inflation. Astron. Astrophys.  {\bf 571}, id.A22 (2014)  [arXiv:1303.5082].

\bibitem[\protect\citeauthoryear{Linde}{2014}]{Linde} A.~Linde. Inflationary Cosmology after Planck 2013.  (2014) [arXiv:1402.0526].


\bibitem[\protect\citeauthoryear{Novosyadlyi}{2011}]{Novosyadlyi} B.~Novosyadlyi, V.~Pelykh, Yu.~Shtanov,  A.~Zhuk. Dark energy and dark matter of the universe: in three volumes. Ed. V. Shulga. -- Vol. 1: Dark matter: Observational evidence and theoretical models (Kiev,  Akademperiodyka, 2013) [arXiv:1502.04177].


\bibitem[\protect\citeauthoryear{Fisher}{1948}]{Fisher} I.Z.~Fisher.
Scalar mesostatic field with regard for gravitational effects.
Zh. Exp. Theor. Phys.,  {\bf 18}, 636-640 (1948) [arXiv:gr-qc/9911008].


\bibitem[\protect\citeauthoryear{Janis}{1968}]{Janis}
A.I.~Janis, E.T.~Newman, J.~Winicour. Reality of the Schwarzschild
singularity. Phys. Rev. Lett.,  {\bf 20}, 878-880 (1968).


\bibitem[\protect\citeauthoryear{Solovyev}{2012}]{Solovyev}
 D.~Solovyev, A.~Tsirulev. General properties and exact models of static self-gravitating scalar field configurations. Classic. Quant. Grav., {\bf 29}, id.055013 (2012).

\bibitem[\protect\citeauthoryear{Stuchlik}{2010}]{Stuchlik}
Z.~Stuchl\'ik, J.~Schee. Appearance of Keplerian discs orbiting Kerr superspinars.
Classic. Quant. Grav., {\bf 27}, id. 215017 (2010) [arXiv:1101.3569].

\bibitem[\protect\citeauthoryear{Chowdhury}{2012}]{Chowdhury}
A.N.~Chowdhury, M.~Patil, D.~Malafarina, P.S.~Joshi.  Circular geodesics
and accretion disks in the Janis-Newman-Winicour and gamma metric spacetimes.
Phys. Rev. D,  {\bf 85}, id. 104031 (2012) [arXiv:1112.2522].


\bibitem[\protect\citeauthoryear{Vieira}{2014}]{Vieira}
R.S.S.~Vieira, J.~Schee, W.~Klu\'zniak, Z.~Stuchl\'ik,  M.~Abramowicz. Circular geodesics of naked singularities in the Kehagias-Sfetsos metric of Horava's gravity
Phys. Rev. D, {\bf 99} id.024035 (2014) [arXiv:1311.5820].


\bibitem[\protect\citeauthoryear{Pugliese}{2011}]{Pugliese}
D.~Pugliese, H.~Quevedo, R.~Ruffini.
Equatorial circular motion in Kerr spacetime.
Phys.Rev.D, {\bf 84} id.044030 (2011) [arXiv:1105.2959].


\bibitem[\protect\citeauthoryear{Pugliese2}{2013}]{Pugliese2}
D.~Pugliese, H.~Quevedo, R.~Ruffini.
Equatorial circular motion in Kerr spacetime.
Phys. Rev. D, {\bf 88}, id. 024042 (2013) [arXiv:1303.6250].

\bibitem[\protect\citeauthoryear{Pugliese3}{2013}]{Pugliese3}
D.~Pugliese, H.~Quevedo, R.~Ruffini.
General classification of charged test particle circular orbits in Reissner--Nordström spacetime
European Phys. J. C, {\bf 77}  id.206 (2017)
[arXiv:1304.2940].

\bibitem[\protect\citeauthoryear{Boshkayev}{2016}]{Boshkayev}
K.~Boshkayev, E.~Gasperin, A.C.~Gutierrez-Pineres, H.~Quevedo, S.~Toktarbay. Motion of test particles in the field of a naked singularity. Phys. Rev. D , {\bf 93}, id.024024 (2016) [arXiv:1509.03827].

\bibitem[\protect\citeauthoryear{Novikov-Thorne}{1974}]{Novikov-Thorne}
I.D.~Novikov,  K.S.~Thorne, 1974, Astrophysics of black holes. In: Black holes (Les astres occlus), p. 343 - 450 (1973).

\bibitem[\protect\citeauthoryear{Reynolds-Nowak}{2003}]{Reynolds-Nowak} C.S.Reynolds, M.A.Nowak. Fluorescent iron lines as a probe of astrophysical black hole systems. Phys.Rept. {\bf 377}  389-466 (2003) [arXiv:astro-ph/0212065].


\bibitem[\protect\citeauthoryear{Lasota}{2014}]{Lasota} J.-P.Lasota. Black Hole Accretion Discs. In: Astrophysics of Black Holes, ed. C. Bambi. Springer, 2016.


\bibitem[\protect\citeauthoryear{Guilbert}{1988}]{Guilbert}
P.W.~Guilbert, M.J.~Rees.  ``Cold'' material in non-thermal sources.
Mon. Notic. Roy. Astron. Soc. {\bf 233}, 475-484 (1988).


\bibitem[\protect\citeauthoryear{Lightman}{1988}]{Lightman}
A.P.~Lightman, T.R.~White. Effects of cold matter in active galactic
nuclei - A broad hump in the X-ray spectra. Astrophys. J.  {\bf 335}, 57-66 (1988).

\bibitem[\protect\citeauthoryear{Fabian}{1989}]{Fabian}
A.C.~Fabian, M.J.~Rees, L.~Stella, N.E.~White. X-ray fluorescence from the inner disc in Cygnus X-1 // Mon. Notic. Roy. Astron. Soc. {\bf 238}, 729-736 (1989).

\bibitem[\protect\citeauthoryear{Bronnikov}{2001}]{Bronnikov}  K.A.~Bronnikov. Spherically symmetric false vacuum: no-go theorems and global structure.	Phys.Rev. D {\bf 64}, id.064013 (2001) [arXiv:gr-qc/0104092].


\bibitem[\protect\citeauthoryear{BronShikin}{2002}]{BronShikin}
K.A.~Bronnikov , G.N.~Shikin.  Spherically Symmetric Scalar Vacuum: No-Go Theorems, Black Holes and Solitons. Gravitation Cosmol. {\bf 8}, 107--116 (2002) [arXiv:gr-qc/0109027].

\bibitem[\protect\citeauthoryear{BronnikovFabris}{2005}]{BronnikovFabris} K.A.~Bronnikov. J.C.~Fabris. Regular phantom black holes. Phys. Rev. Lett. {\bf 96}, id.251101 (2006) [arXiv:gr-qc/0511109].


\bibitem[\protect\citeauthoryear{Nikonov}{2008}]{Nikonov}
V.V.~Nikonov, Ju.V.~Tchemarina, A.N.~Tsirulev.  A two-parameter family
of~exact asymptotically flat solutions to the Einstein-scalar field
equations. Classic. Quant. Grav.. {\bf 25}, id. 138001 (2008).

\bibitem[\protect\citeauthoryear{Azreg2009}{2009}]{Azreg2009}
M.~Azreg-A\"{i}nou, Selection criteria for two-parameter solutions to scalar-tensor gravity. General Relativity Gravit. {\bf 42}, 1427-1456 (2010) [arXiv:0912.1722].

\bibitem[\protect\citeauthoryear{Cadoni2015}{2015}]{Cadoni2015}
M.~Cadoni, E.~Franziny, Asymptotically flat black holes sourced by a massless scalar field.  	 Phys. Rev. D  {\bf 91}, id. 104011 (2015) [arXiv:1503.04734].


\bibitem[\protect\citeauthoryear{Bekenstein}{2002}]{Bekenstein} 	J.D.~Bekenstein,  Black Holes: Classical Properties,  Thermodynamics and Heuristic Quantization. ArXiv:gr-qc/9808028 (1998).

\bibitem[\protect\citeauthoryear{Asanov}{1974}]{Asanov}
R.A.~Asanov, Point source of massive scalar field in
gravitational theory. Theoretical and mathematical physics, {\bf 20}, 66–-70 (1974).

\bibitem[\protect\citeauthoryear{Vasylenko}{2015}]{Vasylenko}
A.A.~Vasylenko, E.V.~Fedorova, B.I.~Hnatyk, V.I.~Zhdanov. Evidence
for a binary black hole in active nucleus of NGC 1194 galaxy? Kinemat. Phys. Celest. Bodies. {\bf 31}, 13--18 (2015).

\bibitem[\protect\citeauthoryear{Fedorova}{2016}]{Fedorova}
E.~Fedorova, A.~Vasylenko, B.I.~Hnatyk, V.I.~Zhdanov. The peculiar
megamaser AGN NGC 1194: Comparison with the warped disk candidates NGC 1068 and NGC 4258. Astronom. Nachr. {\bf 337}, 96-100 (2016).



\bibitem[\protect\citeauthoryear{Bambi}{2013}]{Bambi}
C.~Bambi, D.~Malafarina. K$\alpha$ iron line profile from accretion disks around regular and singular exotic compact objects. Phys. Rev. D  {\bf 88}, id. 064022 (2013).

\end{thebibliography}
\end{document}